# Deterministic Switching of the Néel Vector by Asymmetric Spin Torque

Shui-Sen Zhang,[1,2,3] Zi-An Wang,[3,2] Bo Li,[4,*] Wen-Jian Lu,[3] Mingliang Tian,[1,5] Yu-Ping Sun,[1,3,6] Haifeng Du,[1,7,†] and Ding-Fu Shao[3,‡]

[1] *Anhui Province Key Laboratory of Low-Energy Quantum Materials and Devices, High Magnetic Field Laboratory, HFIPS, Chinese Academy of Sciences, Hefei, Anhui 230031, China*

[2] *University of Science and Technology of China, Hefei 230026, China*

[3] *Key Laboratory of Materials Physics, Institute of Solid State Physics, HFIPS, Chinese Academy of Sciences, Hefei 230031, China*

[4] *MOE Key Laboratory for Nonequilibrium Synthesis and Modulation of Condensed Matter, Shaanxi Province Key Laboratory of Quantum Information and Quantum Optoelectronic Devices, School of Physics, Xi'an Jiaotong University, Xi'an 710049, China*

[5] *School of Physics and Optoelectronic Engineering, Anhui University, Hefei 230601, China*

[6] *Collaborative Innovation Center of Microstructures, Nanjing University, Nanjing 210093, China*

[7] *Institute of Physical Science and Information Technology, Anhui University, Hefei 230601, China*

[*] libphysics@xjtu.edu.cn; [†] duhf@hmfl.ac.cn; [‡] dfshao@issp.ac.cn

Néel vector, the order parameter of collinear antiferromagnets, serves as a state variable in associated antiferromagnetic (AFM) spintronic devices to encode information. A deterministic switching of Néel vector is crucial for the write-in operation, which, however, remains a challenging problem in AFM spintronics. Here we demonstrate, based on analytical derivation and macro-spin simulations, that Néel vector switching can be generally achieved via a current-induced spin torque, provided the spin accumulations responsible for this torque are non-identical between opposite sublattices. This condition occurs widely in AFM films, as symmetry equivalence between sublattice-dependent spin accumulations is usually absent, allowing unequal spin accumulations induced by Edelstein effect or a spin current. Unlike previously studied spin torques induced by uniform or staggered spin accumulations—where either the field-like or damping-like component dominates exclusively—the asymmetric spin torque features cooperative contributions from both components, leading to Néel vector dynamics that are fundamentally distinct from previous expectations. Crucially, the static states stabilized by the asymmetric spin torque enable versatile Néel vector switching strategies—field-free spin-transfer torque switching during current application, as well as field-free or field-assisted spin–orbit torque switching after the current pulse—demonstrating that established spin torque techniques from ferromagnetic spintronics can be directly adapted to AFM systems, a capability absent in previous theoretical frameworks. Our work establishes a general mechanism for current-induced Néel vector switching, which is in principle feasible for all collinear antiferromagnets, and thus paves the route to realize efficient writing in AFM spintronics.

Spintronics exploits magnetic order parameters as state variables to encode information [1]. Current efforts in the field focus on optimizing device operation speed, power consumption, and scalability. Antiferromagnetic (AFM) materials are outstanding candidates for these goals [2,3], owing to their ultrafast magnetic dynamics [4] and negligible stray fields. AFM-based devices—particularly AFM tunnel junctions (AFMTJs) [5]—are thus highly promising for high-speed, energy-efficient, and high-density applications. While the tunneling magnetoresistance (TMR) effect has been theoretically proposed for both collinear [6-9] and noncollinear [10,11] AFMTJs, it has so far been experimentally observed only in AFMTJs based on noncollinear [12-15] or van der Waals antiferromagnets [16,17]. This is largely attributed to the relative ease of manipulating these materials via magnetic fields [16-19] or spin torques [20-32].

In contrast, controlling the Néel vector in collinear antiferromagnets remains challenging. Field-like (FL) spin torques—driven by staggered spin accumulations via the Edelstein effect—have been proposed to switch in noncentrosymmetric antiferromagnets [33-36]. In most other antiferromagnets, however, spin accumulation is expected to be uniform, generating damping-like (DL) torques that induce persistent ultrafast oscillations [37-42]. Although deterministic switching is theoretically possible via DL torques with tailored current pulses [40], such torques have so far only been shown to rotate the Néel vector between easy axes [43-46], unless additional symmetry breaking is present [47,48].

It appears, however, that spin accumulations on two magnetic sublattices are not necessarily identical due to broken symmetry constraints in AFM films. As we demonstrate in this work, based on rigorous analytical derivation and comprehensive macro-spin simulations, such disbalance in spin accumulations can introduce an asymmetric spin torque in an antiferromagnet. Unlike previously studied spin torques induced by uniform or staggered spin accumulations—where either the FL or DL component dominates exclusively—the asymmetric spin torque features cooperative contributions from both components. Consequently, it drives Néel vector dynamics that are fundamentally distinct from previous expectations, offering promising opportunities for Néel vector switching and thus the deterministic control of AFM spintronic devices.

There are two major mechanisms to generate spin accumulations $\delta\boldsymbol{S}_i$ on sublattices $i = A, B$ in a collinear antiferromagnet. The first mechanism is based on Edelstein effect due to space inversion $\hat{P}$ symmetry breaking [23,33], where equal spin accumulations are commonly expected, i.e. $\eta\delta\boldsymbol{S}_A = \delta\boldsymbol{S}_B$ with the asymmetry factor $\eta = 1$. However, the only constraint enforcing $\eta = 1$ is $\hat{T}\hat{t}$ symmetry combining time reversal $\hat{T}$ and fractional lattice translation $\hat{t}$ [23, 49 ]. For noncentrosymmetric antiferromagnets without $\hat{T}\hat{t}$ symmetry, a combined $\hat{P}\hat{T}$ symmetry enforces $\eta = -1$ [33]. Otherwise, there are no symmetry restrictions to exclude $|\eta| \neq 1$ induced by Edelstein effect, especially in AFM thin films where the bulk symmetry is strongly reduced.

The second mechanism originates from spin accumulations $\delta\boldsymbol{S}_i$ induced by an external spin current $J_s$ (via quantum tunneling in AFMTJs or the spin Hall effect), which are determined by the sublattice-dependent attenuation of $J_s$ as it enters the antiferromagnet. Because the interface breaks bulk symmetry, the two sublattices typically absorb different amounts of $J_s$—unless a mirror plane or rotation axis normal to the interface is preserved to connect the atomic sites between the sublattices [50]. For example, in a film with a standard G-type AFM order, sublattice magnetizations, $\boldsymbol{m}_A$ and $\boldsymbol{m}_B$ are alternating within the layers parallel to the interface (Fig. 1(a)). With a flat interface, the average distances of the atoms relative to the interface are identical by symmetry for the two sublattices, leading to $\eta = 1$. On the other hand, for an A-type stacking composed of alternating FM layers, no symmetry operations connecting $\boldsymbol{m}_A$ and $\boldsymbol{m}_B$ at interface, leading to $\eta \neq 1$. Since the A-type AFM film can be realized by engineering the film growth direction of most antiferromagnets, the unequal spin accumulations should be very common for AFM films.

Recent calculations suggest that spins can be even solely accumulated on one sublattice in an A-type van der Waals AFM bilayer, representing an extreme case of $\eta = 0$ [9, 51 ]. Alternately, a large $\delta\boldsymbol{S}_i$ in an AFM film occurs if the sublattice $i$ is more conductive along the film-normal direction (Fig. 1(c)). This is typical for the films of altermagnets [52-54] if the rotation axis connecting the two sublattices is not perpendicular to the film. The recently identified X-type AFM stacking represents an extreme case of this prototype, due to its sublattice selective spin-dependent transport property [55].

We note that $\eta$ can be either positive or negative. When $\delta\boldsymbol{S}_i$ arises solely from an external spin current $J_s$, $\eta$ is typically positive. However, when contributions from the Edelstein effect [33,56,57], Néel spin currents [8], and other sources [58-61] coexist with $J_s$, $\eta$ can become negative if the staggered components are sufficiently strong.

Next, assuming unequal spin accumulations on magnetic sublattices of a collinear antiferromagnet with a uniaxial anisotropy along $\boldsymbol{z}$ direction, we demonstrate that the Néel vector switching can be achieved by the asymmetric spin torque, which hosts torque components neither uniform nor staggered on two sublattices [62][63,64,65,66,67,68,69]. Within the macro-spin approximation, the modified Landau-Lifshitz-Gilbert (LLG) equations for two sublattices, where all parameters are scaled with frequency, are [20,70,71]

$$\dot{\boldsymbol{m}}_A = \boldsymbol{m}_A \times (\omega_E \boldsymbol{m}_B - \omega_K m_{z,A}\boldsymbol{z} - \omega_F \boldsymbol{p} + \alpha\dot{\boldsymbol{m}}_A - \omega_D \boldsymbol{m}_A \times \boldsymbol{p}), \quad (1)$$

$$\dot{\boldsymbol{m}}_B = \boldsymbol{m}_B \times (\omega_E \boldsymbol{m}_A - \omega_K m_{z,B}\boldsymbol{z} - \eta\omega_F \boldsymbol{p} + \alpha\dot{\boldsymbol{m}}_B - \eta\omega_D \boldsymbol{m}_B \times \boldsymbol{p}), \quad (2)$$

where $m_{z,i}$ is the z-component of $\boldsymbol{m}_i$, $\omega_E$ and $\omega_K$ describe the AFM exchange interaction and uniaxial anisotropy, $\alpha$ is the damping factor, and $\boldsymbol{p} = (p_x, p_y, p_z)$ is the unit vector of spin polarization of $\delta\boldsymbol{S}_A$. Without loss of generality, we set $|\delta\boldsymbol{S}_A| > |\delta\boldsymbol{S}_B|$, thus $-1 < \eta < 1$. $\omega_F$ and $\omega_D$ describe the FL and DL torques of sublattice A. Since $\eta\delta\boldsymbol{S}_A = \delta\boldsymbol{S}_B$, the FL and DL torques for sublattice B are thus described by $\eta\omega_F$ and $\eta\omega_D$, respectively.

We introduce the normalized Néel vector $\boldsymbol{n} = (\boldsymbol{m}_A - \boldsymbol{m}_B)/2$ and normalized net magnetization $\boldsymbol{m} = (\boldsymbol{m}_A + \boldsymbol{m}_B)/2$ which satisfy $\boldsymbol{n} \cdot \boldsymbol{m} = 0$ and $\boldsymbol{m}^2 + \boldsymbol{n}^2 = 1$. In the exchange-dominant regime, where $\omega_E$ is significantly larger than $\omega_K$, $\omega_F$ and $\omega_D$, the Néel vector $\boldsymbol{n}$ can be treated as a rigid vector, thus $|\boldsymbol{m}| \ll |\boldsymbol{n}|$, $\boldsymbol{n}^2 \approx 1$, and $\boldsymbol{n} \cdot \dot{\boldsymbol{n}} = 0$. By neglecting the higher-order terms of $\boldsymbol{m}$, we derive the coupled equations for $\boldsymbol{m}$ and $\boldsymbol{n}$ from Eqs. (1) and (2):

$$\dot{\boldsymbol{m}} = -\omega_K n_z \boldsymbol{n} \times \boldsymbol{z} + \alpha \boldsymbol{n} \times \dot{\boldsymbol{n}} + \boldsymbol{\tau}_{st}(\eta), \quad (3)$$

$$\dot{\boldsymbol{n}} = 2\omega_E \boldsymbol{n} \times \boldsymbol{m} - \omega_K (m_z \boldsymbol{n} + n_z \boldsymbol{m}) \times \boldsymbol{z} + \alpha(\boldsymbol{m} \times \dot{\boldsymbol{n}} + \dot{\boldsymbol{n}} \times \boldsymbol{m}) + \boldsymbol{\tau}_{st}(-\eta), \quad (4)$$

where $\boldsymbol{\tau}_{st}$ describes the asymmetric spin torque as

$$\boldsymbol{\tau}_{st}(\eta) = -\frac{\omega_F}{2}[(1-\eta)\boldsymbol{n} \times \boldsymbol{p} + (1+\eta)\boldsymbol{m} \times \boldsymbol{p}] - \frac{\omega_D}{2}(1+\eta)\boldsymbol{n} \times (\boldsymbol{n} \times \boldsymbol{p}) - \frac{\omega_D}{2}(1-\eta)[\boldsymbol{m} \times (\boldsymbol{n} \times \boldsymbol{p}) + \boldsymbol{n} \times (\boldsymbol{m} \times \boldsymbol{p})], \quad (5)$$

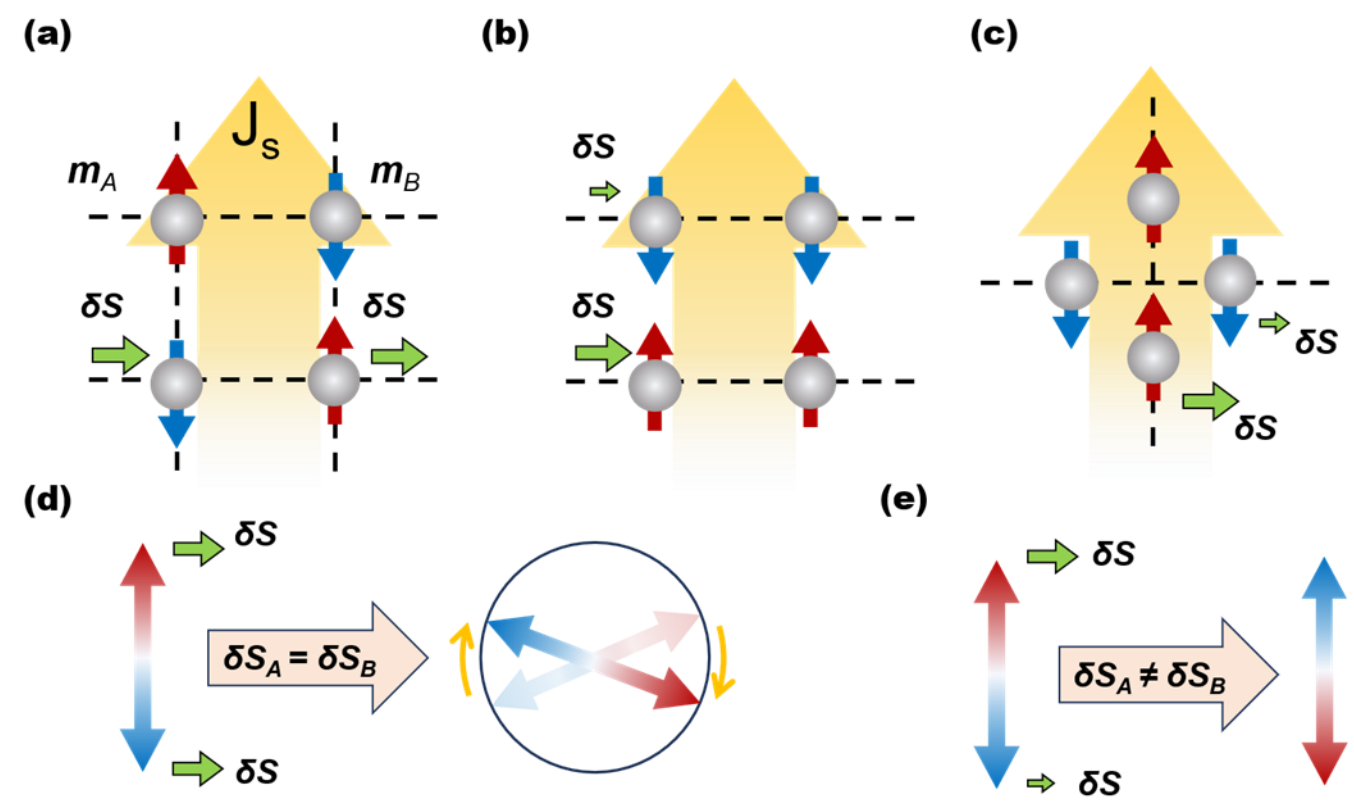

**FIG. 1: Spin accumulations $\delta S$ induced by an external spin current $\boldsymbol{J_s}$ in antiferromagnets.** (**a**) An AFM film where the atomic sites of the two sublattices are symmetric around the film-normal direction, i.e. these sites are connected by symmetry operation such as an in-plane translation and a mirror plane or a rotation axis parallel to the film-normal direction. The $\boldsymbol{J_s}$ along this direction leads to equal spin accumulations on $\boldsymbol{m}_A$ and $\boldsymbol{m}_B$. (**b,c**) AFM films where the atomic sites of the two sublattices are asymmetric around the film-normal direction. The $\boldsymbol{J_s}$ along this direction leads to unequal spin accumulations due to the atomic sites of one sublattice being closer to the spin source (**b**) or being more conductive (**c**). (**d,e**) The effects on Néel vector when applying the equal (**d**) and unequal (**e**) spin accumulations on two sublattices. Here, (**d**) illustrates the previously understood scenario where equal spin accumulations lead to oscillation, while (**e**) depicts the asymmetric case enabling deterministic switching.

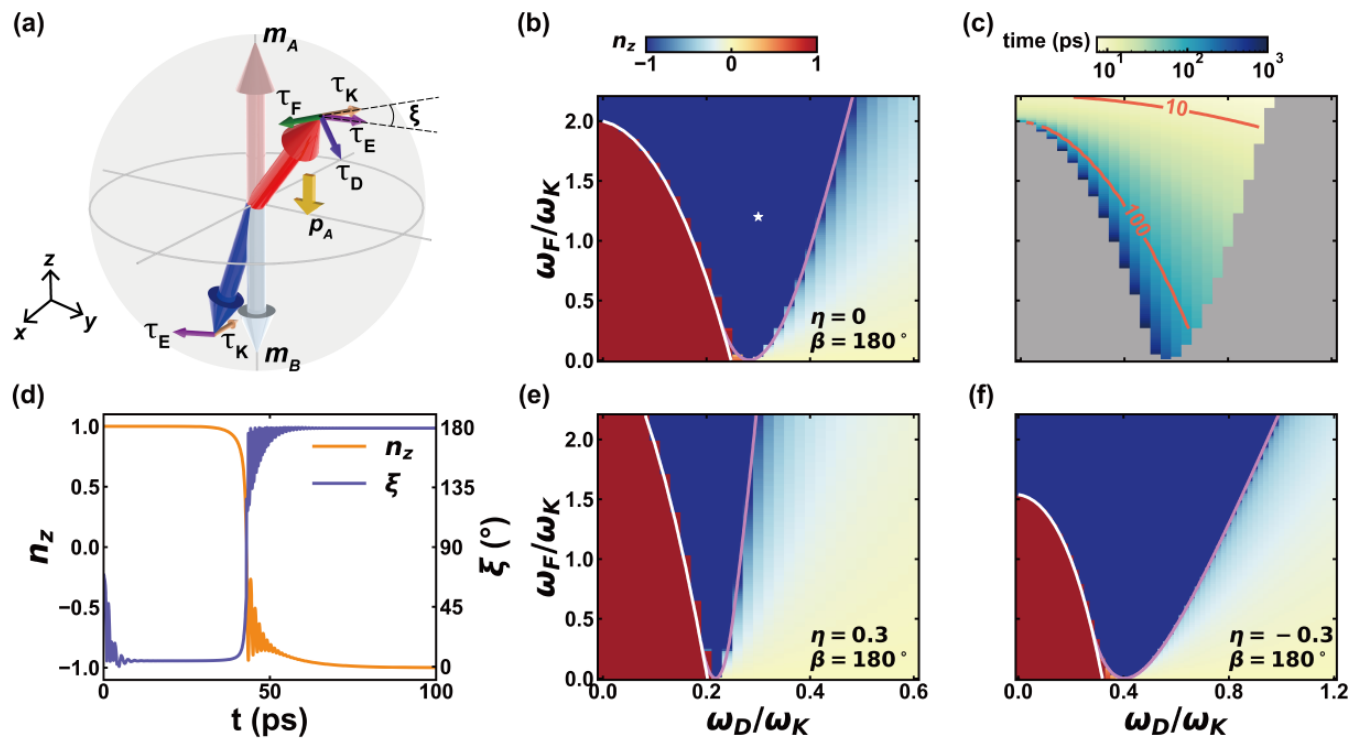

**FIG. 2: Asymmetric spin torque driven by z-polarized unequal spin accumulations. (a)** Magnetic moments ($\boldsymbol{m}_A$: red, $\boldsymbol{m}_B$: blue; light colors: initial orientations) and torques during switching for $\eta = 0$. Small arrows indicate torques driven by anisotropy ($\boldsymbol{\tau}_K$), exchange ($\boldsymbol{\tau}_E$), field-like ($\boldsymbol{\tau}_F$), and damping-like ($\boldsymbol{\tau}_D$); the boxed arrow shows spin polarization $\boldsymbol{p}_A$. $\xi$ is the angle between $\boldsymbol{\tau}_K$ and $\boldsymbol{\tau}_E$. **(b)** Phase diagram of $n_z$ vs. $\omega_D$ and $\omega_F$ ($\eta = 0$) during current application; solid lines mark theoretical switching boundaries. **(c)** Switching time for deterministic switching in **(b)**; gray regions: no switching. **(d)** Time evolution of $n_z$ and $\xi$ for the point marked by the white pentagram in **(b)**. **(e,f)** $n_z$ phase diagrams for $\eta = 0.3$ **(e)** and $\eta = -0.3$ **(f)** during current application.

In Eq. (5), we explicitly retain $\boldsymbol{m}$-dependent terms up to the first order which were conventionally neglected in the earlier works [37-40,72,73]. It can be immediately seen from Eq. (5) that when the infinitesimal terms related to $\boldsymbol{m}$ are neglected, uniform ($\eta = 1$) and staggered ($\eta = -1$) spin accumulations give rise to only the DL and FL components, respectively, consistent with previous derivations [33,37,39]. In contrast, unequal spin accumulations ($-1 < \eta < 1$) lead to the coexistence of both FL and DL components, indicating the magnetic dynamics that are fundamentally different from those previously assumed, as explicitly compared in the Supplemental Material [62].

$\boldsymbol{\tau}_{st}$ induces a weak canting and results in a slave variable $\boldsymbol{m}$ dependent on $\boldsymbol{n}$, as derived from Eq. (4),

$$\boldsymbol{m} = \frac{1}{2\omega_E}\dot{\boldsymbol{n}} \times \boldsymbol{n} + \frac{(1+\eta)\omega_F}{4\omega_E}\boldsymbol{n} \times \boldsymbol{p} \times \boldsymbol{n} + \frac{(1-\eta)\omega_D}{4\omega_E}\boldsymbol{n} \times \boldsymbol{p}, \quad (6)$$

Based on the Lagrange formalism [74,75], the corresponding Lagrangian function $L$ and the dissipative Rayleigh function $R$ are

$$L = \frac{1}{2\omega_E}\dot{\boldsymbol{n}}^2 + (1-\eta)\omega_F \boldsymbol{n} \cdot \boldsymbol{p} + \omega_K \boldsymbol{n}_z^2 + \frac{(1+\eta)\omega_F}{2\omega_E}\boldsymbol{p} \times \dot{\boldsymbol{n}} \cdot \boldsymbol{n}, \quad (7)$$

$$R = \alpha \dot{\boldsymbol{n}}^2 - (1+\eta)\omega_D \boldsymbol{n} \times \boldsymbol{p} \cdot \dot{\boldsymbol{n}} + \frac{(1-\eta)\omega_D}{2\omega_E}(\boldsymbol{p} \cdot \boldsymbol{n})\dot{\boldsymbol{n}}^2. \quad (8)$$

Eqs. (7,8) are the central results of this work, which formalize the Lagrangian of the system with a general asymmetric ratio $\eta$. For $\eta = 1$, $R$ reduces to the typical form, while an additional FL-torque term persists in $L$ compared to the prior derivations [37]. When $\eta \neq 1$, the additional asymmetric terms appear both in $L$ and $R$. Based on Eqs. (7,8), we derived the stability conditions of the stationary solutions [62,76,77].

We consider spin polarization $\boldsymbol{p} = (0, \sin\beta, \cos\beta)$ within the $yz$-plane. When $\beta = \pi$, the formulas above describe the magnetic dynamics induced by $z$-polarized spin accumulation, which is usually associated with a spin-transfer torque (STT) in a device with current perpendicular-to-plane (CPP) structure. This configuration exhibits rotational symmetry about the $z$-axis, resulting in three stationary solutions [62]: 1) $n_z = 1$, indicating that the Néel vector maintains its initial orientation; 2) $n_z = -1$, indicating that the Néel vector is fully reversed by the asymmetric torque; and 3) a constant oscillation around $z$ axis. The second solution is required for the write-in operation in AFM spintronics.

In order to understand and verify these analytical results, we simulate the associated Néel vector dynamics under an asymmetric spin torque induced by a constant current based on the numerical simulations directly using the full sublattice equations Eqs. (1,2) without any other approximation. We take $\omega_K/\gamma = 1$ T, $\omega_E = 100\omega_K$, and $\alpha = 0.01$, where $\gamma$ is the gyromagnetic ratio, and conclude the average $n_z$ within a short time window around $t = 1$ ns. Figure 2 shows the cases for $\eta = 0$, 0.3, and −0.3. We find that in the simulated phase diagrams of $n_z$ with respect to different $\omega_D$ and $\omega_F$ (Fig. 2(b,e,f)), there are typically three regions in the phase diagram, i.e. $n_z = 1$ (the red region) indicating the Néel vector is maintained along the initial direction, $n_z = -1$ (the blue region) indicating the Néel vector is fully reversed from the initial state, and $-1 < n_z < 1$ (the region with lighter colors) for a constant oscillation. We find our simulation perfectly consistent with the analytical boundaries (the white and purple lines in Fig. 2(b,e,f)). We note the area of the region of $n_z = -1$ associated with the deterministic Néel vector switching is smaller for the positive $\eta$ and larger for the negative $\eta$.

For simplicity, we consider the $\eta = 0$ case (Fig. 2(a)) to illustrate the Néel vector dynamics. When the FL ($\boldsymbol{\tau}_F = -\omega_F \boldsymbol{m}_A \times \boldsymbol{p}$) and DL ($\boldsymbol{\tau}_D = -\omega_D \boldsymbol{m}_A \times (\boldsymbol{m}_A \times \boldsymbol{p})$) torques on $\boldsymbol{m}_A$ overcome the anisotropy torque ($\boldsymbol{\tau}_{K,A} = -\boldsymbol{m}_A \times \omega_K m_{z,A}\hat{\boldsymbol{z}}$), a canting between $\boldsymbol{m}_A$ and $\boldsymbol{m}_B$ develops, producing a nonzero exchange torque $\boldsymbol{\tau}_{E,A} = \omega_E \boldsymbol{m}_A \times \boldsymbol{m}_B$. For small canting, $\boldsymbol{\tau}_{E,A}$ is nearly parallel to $\boldsymbol{\tau}_{K,A}$, driving ultrafast precession about the *z*-axis. Meanwhile, a small component of $\boldsymbol{\tau}_{E,A}$ perpendicular to $\boldsymbol{\tau}_{K,A}$, together with $\boldsymbol{\tau}_D$, gradually pull the Néel vector toward the opposite direction and amplifies the canting. Once the canting is large enough, a finite angle $\xi$ emerges between $\boldsymbol{\tau}_{K,A}$ and $\boldsymbol{\tau}_{E,A}$(Fig. 2(a)), triggering rapid switching. Near the first phase boundary in the phase diagram, this process is slow (Fig. 2(c,d)), but stronger $\boldsymbol{\tau}_F$ and $\boldsymbol{\tau}_D$ reduce the switching time to $\sim 10$ ps. Beyond the second boundary, large $\omega_D$ and thus $\boldsymbol{\tau}_D$ causes compensation of the $\theta$-components of the torques, which hinders the switching and thus leads to the constant oscillation of Néel vector.

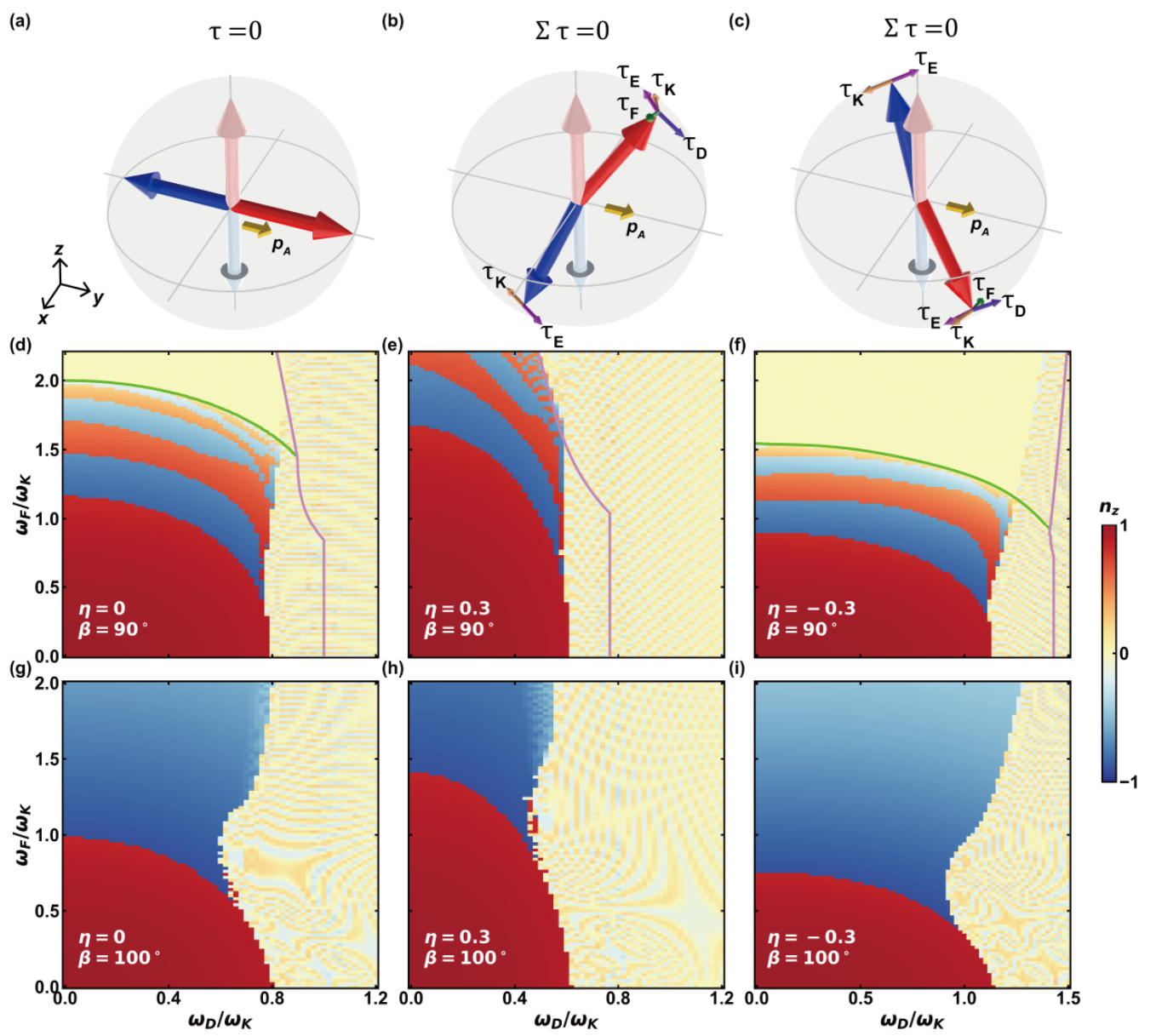

**FIG. 3: Asymmetric spin torque from unequal spin accumulations polarized in the *y*-*z* plane.** (**a–c**) Schematics of magnetic moments and torques for static states with $n_z = 0$ (**a**), $n_z > 0$ (**b**), and $n_z < 0$ (**c**), for $\eta = 0$ and spin polarization $\boldsymbol{p} \parallel \hat{y}$. (**d–f**) Phase diagrams of $n_z$ under current for $\boldsymbol{p} \parallel \hat{y}$ and $\eta = 0$ **(d)**, 0.3 **(e)**, and $-0.3$ **(f)** during current application. Solid lines denote theoretical boundaries of static solutions: $n_z \neq 0$ in the region covered by the green line; $n_z = 0$ between green and purple lines. (**g–i**) Corresponding phase diagrams for tilted polarization with $p_z/p_y = -\tan 10° = -0.176$ and $\eta = 0$ (**g**), 0.3 (**h**), $-0.3$ (**i**).

We also consider the case with $\boldsymbol{p} = (0, \sin\beta, \cos\beta)$ with $\beta = \frac{\pi}{2}$, which represents the asymmetric torque due to unequal *y*-polarized spin accumulations by conventional spin-orbit coupling (SOC) effects such as SHE and Edelstein effect in a device with current in-plane (CIP) geometry. There are two static solutions. The first type is $n_z = 0$ where the Néel vector is reoriented exactly along *y* direction, since all torques in Eqs. (1,2) vanish for this state (Fig. 3(a)). This state is unstable once the spin torque is removed and relaxes with equal probability to the positive or negative *z* direction, and therefore cannot enable deterministic switching. We find that one of the constraints for the state of $n_z = 0$ is $\omega_F > \omega_K/(1-\eta)$, indicating this state is more difficult to achieve for a positive $\eta$. On the other hand, since this configuration exhibits a two-fold rotational symmetry about the *y*-axis, the second solution (with $n_z \neq 0$) consists of two degenerate states [62], i.e. $0 < n_z < 1$ and $-1 < n_z < 0$ (Fig. 3(b,c)), provided that all the torques in Eqs. (1,2) are compensated (Fig. 3(b,c)). When the asymmetric spin accumulations are released, the state with $n_z > 0$ returns to the initial state, while the state with $n_z < 0$ results in the deterministic switching. Note there is no solution of constant oscillation, since the anisotropy torque changes during the oscillation.

Our macro-spin approximation clearly proves the existence of the three states. As shown in the phase diagrams for $\eta = 0$, 0.3, and $-0.3$ (Fig. 3(d-f)), the state of $n_z = 0$ occurs only for a large $\omega_F$, which is relatively easier to achieve when $\eta$ is negative, as expected by our analytical derivation. On the other hand, the states of $0 < n_z < 1$ and $-1 < n_z < 0$ emerge alternately for smaller $\omega_F$, implying these two states are indeed degenerate and thus can be easily swapped. The analytical boundaries (the green and purple lines) accurately separate the state of $n_z = 0$ and other static states obtained by simulation. On the other hand, we find that the analytical boundaries cover some regions of the Néel vector oscillation, possibly because of the emergent nonstationary state in nonlinear dynamic [62]. This does not qualitatively influence our conclusion about the Néel vector switching.

Since $\boldsymbol{\tau}_F$ and $\boldsymbol{\tau}_D$ generated by the *y*-polarized spin accumulations do not compete with $\boldsymbol{\tau}_{E,A}$ in the initial state, the time required to make $n_z$ remaining negative is typically below 10 ps, promising for the ultrafast applications. However, in order to realize the deterministic switching of the Néel vector within the narrow regions of the negative $n_z$ in Fig. 3(d-f), a subtle tuning of $\omega_F$ and $\omega_D$ is required, which can be controlled by turning the current density. We find that this requirement can even be eliminated by introducing the unequal spin accumulations with a tilted spin polarization $\boldsymbol{p} = (0, \sin\beta, \cos\beta)$, where $\beta = \frac{\pi}{2} + \beta'$ results in $p_z/p_y = -\tan\beta' < 0$. Such an unconventional spin polarization can be realized by using the low-symmetry spin sources, which has been widely demonstrated experimentally [78]. As shown in Fig. 3(g-i), the state with negative $n_z$ become more stable in the presence of a small $\beta'$, desirable for the deterministic switching in realistic devices.

The existence of a state with $n_z = 0$, where the Néel vector is aligned exactly along the *y*-direction, resembles the situation of conventional spin-orbit torque (SOT) in a perpendicularly magnetized ferromagnet driven by a *y*-polarized spin accumulation [23]. In ferromagnets, deterministic switching can be achieved with an *x*-directional assisted magnetic field ($\boldsymbol{H}_x$)

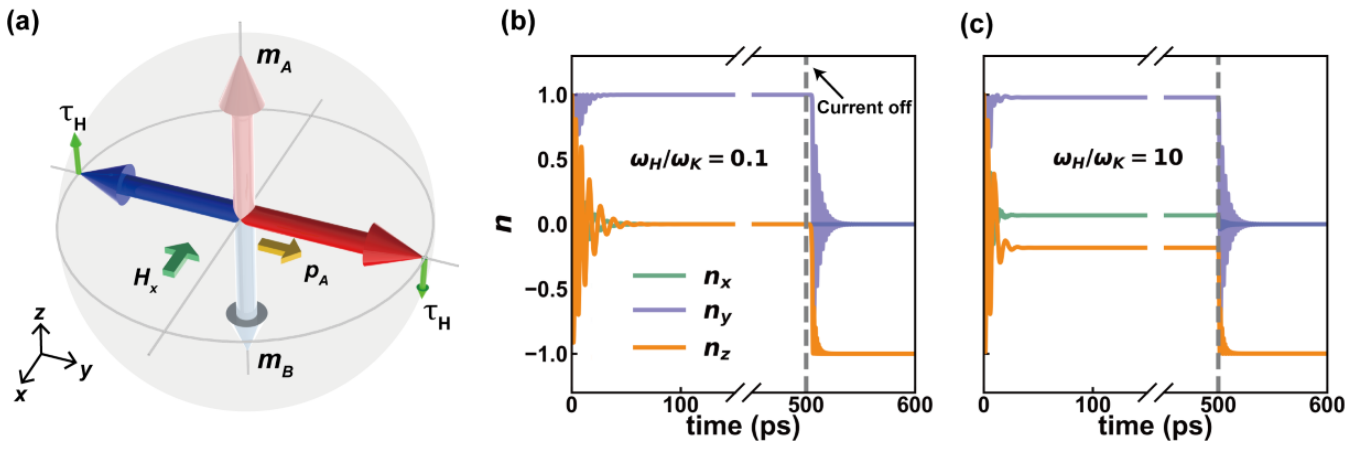

**FIG. 4: Field-assisted deterministic Néel vector switching. (a)** Schematic of staggered torques ($\tau_H$, small green arrows) induced by an in-plane field $H_x$ (green boxed arrow), which enables deterministic control of the final switching direction via the static state where the Néel vector is along $y$-direction. (**b,c**) Switching dynamics for $\eta = 0$ under constant $H_x = \omega_H/\gamma$ with $\omega_H/\omega_K = 0.1$ (**b**) and 10 (**c**). To emulate an experimentally realistic short current pulse, $\omega_F = 1.9\omega_K$ and $\omega_D = 0.6\omega_K$ are applied for the first 500 ps and then turned off. The switching remains robust even for significantly shorter or longer pulse durations.

[23]. We find that this can also be employed to realize deterministic switching in the $n_z = 0$ regime of antiferromagnets. As shown in Fig. 4(a), when the Néel vector is oriented along the $y$-direction, an $\boldsymbol{H}_x$ generates staggered torques of the form $\sim\boldsymbol{m}_i \times \boldsymbol{H}_x$ ($i = A, B$). These torques induce a finite $n_z$, enabling deterministic control over the final switching direction by selecting the sign of $\boldsymbol{H}_x$ and thus of $n_z$.

Figure 4(b) demonstrates that, in the presence of a small $\boldsymbol{H}_x$ (with $|\boldsymbol{H}_x| = \omega_H/\gamma$ and $\omega_H/\omega_K = 0.1$), deterministic Néel vector switching is achieved immediately after the pulse, despite the initially vanishingly small $n_z$. Notably, this switching remains robust even for $\boldsymbol{H}_x$ significantly exceeding the anisotropy field, as demonstrated for $\omega_H/\omega_K = 10$ (Fig. 4(c)). Such large fields would be prohibitive in ferromagnets, where they would reorient the magnetization toward the $x$-direction. In antiferromagnets, however, the strong exchange coupling suppresses such a reorientation and only allows a canting with moderate $n_z$ without qualitatively altering the switching dynamics. This mechanism provides a natural explanation for recent experiments of field-assisted SOT switching in A-type antiferromagnets [79,80], where robust switching persists even under a very large $|\boldsymbol{H}_x| = 3$ T [79].

Although Néel-vector switching in antiferromagnets with multi-axial magnetic anisotropy lies beyond the primary scope of this study, our proposed mechanism is fully applicable to such systems. Crucially, our static-state-based approach provides the deterministic control over the final Néel-vector orientation, enabling, for example, reliable selection between $+x$ and $-x$ (or $+y$ and $-y$) in materials with orthogonal in-plane easy axes—a level of precision that is typically unattainable with oscillation-based switching schemes.

Moreover, we point out that the asymmetric spin torque naturally exists in bulk AFM systems with lower symmetry, such as the half-metallic antiferromagnets [81,82], or ferrimagnets at the compensation temperature [83].

In conclusion, we have demonstrated that the deterministic switching of the Néel vector in collinear AFM films can be driven by an asymmetric spin torque, arising from the previously overlooked disbalance of spin accumulations on the two magnetic sublattices which is typical for most antiferromagnets. Crucially, the static states stabilized by the asymmetric spin torque enable versatile Néel vector switching strategies—field-free STT switching during current application, as well as field-free or field-assisted SOT switching after the current pulse—demonstrating that established spin torque techniques from ferromagnetic spintronics can be directly adapted to antiferromagnetic systems. Overall, our work establishes a general mechanism for realizing current-induced Néel vector switching which is in principle feasible for all collinear AFM systems, thereby paving the way for efficient writing in AFM spintronics.

**Acknowledgments.** We thank Evgeny Y. Tsymbal, Pietro Gambardella, Wenqing He, Yumeng Yang, Xuepeng Qiu, Lan Wang, and Zhaochu Luo for the helpful discussions. This work was supported by the National Key R&D Program of China (Grant No. 2024YFB3614100), the National Natural Science Funds for Distinguished Young Scholar (Grant No. 52325105), the National Natural Science Foundation of China (Grants Nos. 12241405, 12274411, and 12404185), the Basic Research Program of the Chinese Academy of Sciences Based on Major Scientific Infrastructures (Grant No. JZHKYPT-2021-08), and the CAS Project for Young Scientists in Basic Research (Grant No. YSBR-084). Simulations were performed at Hefei Advanced Computing Center.